\documentclass[twocolumn,showpacs,amsmath,amssymb,prc,floatfix]{revtex4}% Phys. Rev.

\usepackage{graphicx}% Include figure files
\usepackage{dcolumn}% Align table columns on decimal point
\usepackage{bm}% bold math

\begin{document}

%\preprint{APS/123-QED}

\title{Improved Nearside-Farside Method for Elastic 
Scattering Amplitudes}

\author{R. Anni}
\email{Raimondo.Anni@le.infn.it}
\affiliation{Dipartimento di Fisica dell'Universit\`a and
Istituto Nazionale di Fisica Nucleare, I73100 Lecce, Italy}

\author{J. N. L. Connor}
\email{J.N.L.Connor@Manchester.ac.uk}
\author{C. Noli}
\affiliation{Department of Chemistry, University of Manchester, 
Manchester M13 9PL, United Kingdom}

\date{July 18, 2002} % It is always \today, today,

\begin{abstract}
A simple technique is described that provides an improved 
nearside-farside (NF) method for elastic scattering 
amplitudes.
The technique, involving the novel resummation of a Legendre partial
wave series, reduces the importance of unphysical contributions to 
NF subamplitudes, which can arise in more conventional NF decompositions.
Detailed applications are made to a strong absorption model and to a 
$^{16}$O + $^{16}$O optical potential at $E_{\text{lab}} = 145$ MeV.
We also discuss
$^{16}$O + $^{16}$O at $E_{\text {lab}}$ = 480, 704, 1120 MeV, 
and $\alpha$ + $^{12}$C, 
$\alpha$ + $^{40}$Ca, both at $E_{\text {lab}}$ = 1370 MeV.

\end{abstract}

\pacs{24.10.Ht, 25.70.Bc, 34.50.-s, 03.65.Sq}

\maketitle

%------------------
\section{Introduction}
\label{Sec:1}

In heavy-ion, atomic and molecular collisions, an elastic 
differential cross section, $\sigma(\theta)$, where $\theta$ is
the scattering angle, is often characterized by a
complicated interference pattern.
This complicated structure makes it difficult to understand the  
physical phenomena involved in the scattering process, as well as 
the links between $\sigma(\theta)$ and the properties of the model 
that describes the phenomenon.

In some cases, semiclassical methods \cite{BRI85} explain the scattering
pattern as the interference between simpler, and slowly varying,
subamplitudes.
If we ignore the complication that, in some angular regions, 
uniform asymptotic techniques are necessary, then the 
semiclassical subamplitudes arise mathematically from saddle 
points or poles, which account physically for contributions 
from reflected, refracted or generalized diffracted 
semiclassical trajectories \cite{NUS92}.
The subamplitudes can be conveniently grouped into two types:
those arising from semiclassical trajectories which 
initially move in the same half plane as the detector
(N or nearside trajectories) and those from the opposite half
plane (F or farside trajectories).

Semiclassical methods are not always simple to apply and 
sometimes they have a limited range of applicability. 
Their limitations are determined by the range of validity
of the (presently known) asymptotic techniques that are used
to approximate the original quantum mechanical
problem.

In order to overcome these difficulties, it is common practice  
to apply to the elastic scattering amplitude, $f(\theta)$,
a NF method that was proposed by Fuller\cite{FUL75} more than 25 years ago.
The Fuller NF method has the merit of being simple 
and, although inspired by the semiclassical theories, it
uses only scattering matrix elements $S_l$ calculated 
(or directly parametrized) by exact quantum mechanics.
The NF subamplitudes are obtained by exact
summation of NF partial wave series (PWS), thereby bypassing 
problems associated with the applicability and 
validity of semiclassical techniques, e.g. using 
approximate $S_l$, replacing the PWS 
by integrals, using stationary phase integration, etc.

The Fuller method is based on a splitting of the
Legendre polynomials, $P_l(\cos \theta)$, in the PWS
of $f(\theta)$, into traveling angular 
wave components, with the traveling angular
waves consistent with detailed 
semiclassical analyses
of scattering from impenetrable and
transparent spheres \cite{NUS92}. 
 
The Fuller method was \cite{HUS84,BRA97} and 
continues to be widely used; indeed 
the ISI Web of Science reports about 140 citations since 1981
to Fuller's seminal work (for more recent examples see 
\cite{OGL98,NIC99,KHO00,OGL00,SZI01,KHO02} and references therein).  
The success of the method 
depends, apart from its simplicity, 
on its remarkable capability of physically explaining complicated 
interference patterns in cross sections as arising 
from the interference between NF subamplitudes having simple 
properties.
In particular, the NF cross 
sections are often less structured and more slowly varying 
with $\theta$ than is the full cross section.
Even though no semiclassical technique has been used, these 
NF subamplitudes can often be given a physical 
interpretation 
%\cite{HUS84} 
(in analogy
with results from semiclassical methods) as 
contributions from simple scattering mechanisms, which then allows
a good understanding of the angular scattering.

In the light of these unquestionable successes, it is
desirable to extend the validity of the Fuller
approach to cases where the original Fuller
method is not (physically) satisfactory, for 
example, it may produce oscillatory and rapidly
varying NF cross sections, when the full cross section
is monotonic and slowly varying with $\theta$.
Examples of these shortcomings have been known for a long 
time. 
One classic example is pure Coulomb scattering. 
For repulsive Coulomb potentials only a N contribution is
expected semiclassically (\cite{BRI85}, p. 56), whereas the 
Fuller NF method yields also a F contribution 
\cite{FUL75}.
As a result, the NF cross sections are less simple than
the full one.
In this case, the unsatisfactorily effects are, however,  
confined to a restricted backward angular region (\cite{FUL75}, p. 1564).
Another more striking example is observed in the 
scattering by a uniformly charged sphere 
(\cite{HUS84}, p. 154, Fig. 26). 
In this case, the ratio of 
the full cross section to the Rutherford one decreases
monotonically
into the shadow of the Coulomb rainbow.
In contrast, the N ratio closely follows the full ratio up to 
$\theta \approx 40^\circ$ when it becomes approximately constant
(i.e. independent of $\theta$), being approximately equal to 
the F ratio. 

A similar effect is also observed in the angular 
distributions for a strong absorption model (SAM) 
with a two parameter ($\Lambda$ and $\Delta$) 
symmetric $S$ matrix element and Fermi-like form 
factors \cite{HAT89}.
For a fixed value of the cut-off parameter $\Lambda$ 
and for a sufficiently large value of the diffuseness 
parameter $\Delta$, the Fuller NF cross sections show 
an almost exponential decline up 
to a certain $\theta$ (which decreases with 
increasing $\Delta$). 
At larger angles, the NF cross sections are greater then the 
full cross section, which continues its oscillatory exponential 
decline.

Similar striking effects appear at high energies and for
large scattering angles in the NF cross sections 
of $\alpha$ particles and light heavy-ions scattered 
by nuclei using optical potentials.
Less striking, but still disturbing, effects are also 
observed, at lower energies, in some N cross sections for the 
optical potentials used to fit recent data of light heavy-ion 
scattering \cite{OGL98,NIC99,KHO00,SZI01}. 
Typical N cross sections rapidly decrease from $0^\circ$ 
and from $180^\circ$. 
The two branches meet in the crossing region where an interference 
pattern appears, with strong oscillations over an extended 
range of angles.

In this paper, we show how some of these shortcomings can
be removed using a new NF method
\cite{ANN01d}, based on an {\it improved modified resummation} 
of the PWS. 
The new method is a development of the Hatchell \cite{HAT89} idea
of incorporating the Yennie, Ravenhall and Wilson (YRW) 
\cite{YEN54} resummation technique into the NF formalism.
The limitations of the NF Hatchell resummation technique have been  discussed 
\cite{McC95,HOL99a,HOL99b} and a {\it modified} NF YRW resummation 
procedure, depending on two parameters $\alpha$ and $\beta$, 
was proposed \cite{WHI01,NOL02} to bypass the difficulties with the
original NF YRW approach.
The possibility of further {\it improving} the modified YRW resummation
procedure, using different resummation parameters $\alpha_1,
\alpha_2, \ldots$ and $\beta_1, \beta_2, \ldots$, together
with a rule to fix the value of these parameters \cite{ANN01d}, is 
discussed in the present work.

For all three NF methods, Fuller, Hatchell and 
ours, the starting point is the quantum mechanical 
PWS for the full scattering amplitude 
$f(\theta)$,
\begin{equation}
f(\theta )=\frac{1}{2ik} \sum_{l=0}^{\infty }a_{l}P_{l}(\cos \theta ),
\label{ParDev}
\end{equation}
where $k$ is the wavenumber, $P_{l}(\cos \theta )$ is the Legendre
polynomial of degree $l$ and $a_{l}$ is given in terms of the 
scattering matrix element $S_l$ by:
\begin{equation}
a_l= (2 l+ 1) (S_l - 1).
\label{ParAmp}
\end{equation}
In the following, we will write $x = \cos \theta$.
For future convenience, we recall that the PWS (\ref{ParDev}), 
considered as a distribution, converges to its exact value  
for $\theta \ne 0$,  upon dropping the 1 in the term 
$S_l -1$ on the r.h.s of (\ref{ParAmp}). 
The omitted amplitude is proportional to the Dirac
delta function $\delta(1- {x})$ (e.g. see \cite{BRI85}, p. 52).
We also recall that the PWS (\ref{ParDev}), considered as a 
distribution, is convergent if $S_l$ is asymptotically 
Coulombic \cite{TAY74}.

In Sec. \ref{Sec:2}, we briefly outline the original Fuller
NF method, and show that unsatisfactorily results are obtained 
for seven collision systems, which are different from those 
considered in \cite{ANN01d}. 
In Sec. \ref{Sec:3}, we discuss a modification 
of the Fuller NF technique proposed in 
Refs. \cite{HOL99a,HOL99b,WHI01,NOL02} and present an
improved modified method.
Our new method is very effective in cleaning
unphysical contributions from NF cross sections 
for the seven examples where the usual Fuller technique 
gives usatisfactorily results.
Our conclusions are in Sec. \ref{Sec:4}.

%--------------------------------------
\section{Limitations of the  Fuller NF method}
\label{Sec:2}

%------------------
\subsection{Introduction}
\label{Sec:20}

The Fuller NF decomposition is realized by splitting in 
(\ref{ParDev}) $P_l({x})$, considered  as a standing 
angular wave, into traveling angular wave components
\begin{equation}
P_l({x})=Q_l^{(-)}({x})+Q_l^{(+)}({x}),
\label{EulDec}
\end{equation}
where (for $x \ne \pm 1$)
\begin{equation}  
\label{Qpm}
Q_l^{(\mp)}({x}) = 
\frac {1}{2}[ P_l({x}) \pm \frac {2i}{\pi}
Q_l({x}) ],
\end{equation}
with $Q_l({x})$ the Legendre function of the second 
kind of degree $l $.

Inserting (\ref{EulDec}) into (\ref{ParDev}) splits $f(\theta )$ 
into the sum of two subamplitudes
\begin{equation}  
\label{NFFul}
f(\theta) = f^{(-)}(\theta) + f^{(+)}(\theta),
\end{equation}
with
\begin{equation}
f^{(\mp )}(\theta )=
\frac{1}{2ik} \sum_{l=0}^{\infty }a_{l} Q_l^{(\mp)}({x} ).
\label{ParDevmp}
\end{equation}
Note that, by construction, the decomposition (\ref{NFFul}) is 
{\it exact}. 
Also we will obtain an exact decomposition, by
using in place of the $Q_l^{(\mp)}({x} )$ 
in (\ref{EulDec}), any pair of functions whose sum 
is $P_l({x})$.
The property of the $Q_{l}^{(\mp )}({x} )$ that makes the
splitting (\ref{EulDec}) important, is the asymptotic
result
\begin{equation}
Q_{l}^{(\mp )}({x} )
\sim 
\sqrt{\frac{1}{2\pi \lambda \sin \theta }}
\exp [\mp i(\lambda \theta -\frac{\pi }{4})],  
\label{AsyQpm}
\end{equation}
for $l \sin \theta \gg 1$, where $\lambda =l+\frac{1}{2}$.
In particular, (\ref{AsyQpm}) allows $(-)$ to be identified with
N scattering and $(+)$ with F scattering (\cite{HUS84}, p. 121). 
In the semiclassical theory, the splitting of
$P_{\lambda - \frac{1}{2}}({x} )$ into the sum of 
$Q_{\lambda - \frac{1}{2}}^{(\mp )}({x} )$, 
or the related splitting obtained from the asymptotic expansions
of these functions \cite{BRI85}, 
plays a crucial role in deriving the semiclassical 
subamplitudes.
In particular, the NF semiclassical subamplitudes arise from terms originally
containing $Q_{\lambda - \frac{1}{2}}^{(\mp )}({x} )$,
or their asymptotic expansions (\ref{AsyQpm}).
These facts raise the hope that the direct calculation of the
$f^{(\mp )}(\theta )$  from their PWS representation in 
(\ref{ParDevmp}) will separate the NF contributions to 
$f(\theta )$, thereby avoiding problems
connected with the applicability or validity of 
the semiclassical theory.

In order to make this hope mathematically rigorous, 
one should prove that
it is possible to perform on the PWS, written in terms of the
$Q_l^{(\mp )}({x})$, the same manipulations that are
used in deriving the complete semiclassical decomposition of 
$f(\theta)$.
These manipulations are essentially path deformations 
in $\lambda$ of the integrals into which (\ref{ParDev}) 
can be transformed, using either the Poisson sum formula
(\cite{NUS92}, p. 45)
or the Watson transformation (\cite{NUS92}, p. 49).
The consequences of these path deformations depend on the
properties of the terms in the PWS when they are continued 
to real or complex values of $\lambda $ from the initial 
physical half integer $\lambda$ values.
The splitting of $P_{l}({x} )$ into 
the $Q_{l}^{(\mp)}({x} )$ 
modifies these properties and can cause the appearance 
of unphysical contributions in the $f^{(\mp )}(\theta )$, which cancel out
in $f(\theta)$ (these contributions are not expected 
to be present in the semiclassical subamplitudes).

In spite of these possible limitations, extensive experience 
with the Fuller NF method has demonstrated that the method is 
usually reliable, in the sense that it often
decomposes $f(\theta)$ into simpler NF subamplitudes,
apparently free from unphysical contributions arising from 
the above mathematical difficulties. 
However for a few examples, some of which were mentioned
in Sec. \ref{Sec:1}, the Fuller NF subamplitudes can 
be directly compared with the corresponding exact analytical, or
semiclassical, results and disagreements are
observed.

Fortunately, the Fuller NF subamplitudes  
contain information that  allows one to recognize the 
unphysical nature of the undesired contributions.
Suppose $f^{(+)}(\theta)$, or $f^{(-)}(\theta)$,  
contains a single semiclassical contribution from a stationary 
phase point at $\lambda(\theta)$. 
Then the derivative with respect to  $\theta$ of the phase 
of $f^{(+)}(\theta)$, or $f^{(-)}(\theta)$, is equal to 
$\lambda(\theta)$, or $-\lambda(\theta)$, respectively 
(\cite{BRI85}, p. 57).
Following Fuller we will call this derivative the {\it Local 
Angular Momentum} (LAM) for the F (or N) subamplitude; it 
depends on $\theta$. 
Usually, only for a generalized diffracted trajectory, arising
from a simple pole, is the LAM  expected to be constant, 
being equal to the angular momentum of the incoming particle 
responsible for the diffraction.

In the semiclassical regime, this constant value is expected 
to be large. 
Because of this, if we observe that, in a certain $\theta$ 
range, LAM $\approx$ 0, this can be considered a warning
for the unphysical nature of the N or F subamplitudes 
in that range of $\theta$.
This occurs for the LAM of the Fuller Coulomb F 
subamplitude, and for  the NF subamplitudes of the SAM in 
the angular region where the NF cross sections
contain unphysical contributions \cite{ANN01d}.
In both cases, this decoupling of $\theta$ from LAM,
together with the fact that the full cross section is
simpler then the NF ones, suggests the unphysical nature 
of the NF subamplitudes.

We show below for seven collision systems, different from 
those considered in \cite{ANN01d}, how unphysical 
NF contributions manifest themselves.
The seven collision systems are; 
(a) a simple SAM model, 
(b) $^{16}$O + $^{16}$O at $E_{\text {lab}}$ = 145 MeV, using the 
WS2 optical potential of Ref. \cite{KHO00},
and 
(c) more briefly,
$^{16}$O + $^{16}$O at $E_{\text {lab}}$ = 480, 704, 1120 MeV, 
and $\alpha$ + $^{12}$C,
$\alpha$ + $^{40}$Ca, both at $E_{\text {lab}}$ = 1370 MeV.

%------------------
\subsection{Strong absorption model for elastic scattering}
\label{Sec:21}

The first example is a simple SAM in which the $S_l$ is directly 
parametrized by
\begin{equation}
\label{SAM}
S_l \equiv S(\lambda)=\frac{1}{1+\exp(\frac{\Lambda-\lambda}{\Delta})}+
    \frac{1}{1+\exp(\frac{\Lambda+\lambda}{\Delta})},
\end{equation}
with $\lambda=l+\frac{1}{2}$, $\Lambda =10.0$ and $\Delta=1.8$,

The SAM (with modifications for
the Coulomb interaction, and in a slightly
different form from (\ref{SAM}) ) was widely used
in early studies of heavy-ion elastic scattering \cite{FRA84,GRY01}.
At the present time, the SAM is not so popular,
using either simple forms such as (\ref{SAM}) or more sophisticated 
functions.
It has been found that the characteristics of 
heavy-ion angular distributions, measured over wide angular 
ranges, are not accounted for by the simpler SAM models;
instead the angular scattering is more easily described 
using optical potentials, rather then attempting complicated 
extensions of the SAM.

In spite of this, the SAM in its simple form (\ref{SAM}) 
continues to be of interest, since it allows important
tests of NF decompositions \cite{HAT89,HOL99a,HOL99b,NOL02}.
This is because the PWS for the SAM (\ref{SAM})
can be evaluated easily by saddle point techniques
\cite{HAT89}, or more simply, using the Watson transformation 
and elementary complex integration.
Both methods allow a simple, mathematically correct, 
identification of the NF subamplitudes.
For $\Lambda \gg 1$, $\exp(- 2 \pi \Delta) \ll 1$ and 
$\Lambda \sin \theta \gg 1$, it is found that the 
NF subamplitudes are 
$\propto \exp(- \pi \Delta \theta \mp  i \Lambda \theta)/\sqrt{\sin 
\theta}$,
to a good degree of approximation (Ref. \cite{HUS84}, Eq. (3.5) ).
The NF cross sections, multiplied by 
$\sin{\theta}$ are equal and have an exponential
slope, whereas the phase derivatives of the NF subamplitudes
are expected to be equal to $\mp \Lambda$, respectively.

The results obtained by applying the Fuller NF method 
to the SAM with parameters $\Lambda=10.0$ and $\Delta=1.8$
are shown in Fig. 1.
In the lower panel, we show a log plot of the dimensionless 
quantity $4 k^2 \sigma(\theta) \sin \theta$ versus $\theta$ 
since the corresponding NF quantities are expected to have 
an exponential slope.
This is additionally shown in Fig.1 by the thin 
dot-dash line which represents 
$\log_{10} [\exp (-2 \pi \Delta \theta)]$.
Furthermore, because the $S_l$ are real, $f(\theta)$ has a
constant phase (and its phase derivative is of no interest), 
while the $f^{(\mp)}(\theta)$ have identical moduli but 
opposite phases. 
Thus we need only show the N (or F) LAM and similarly 
for the cross sections.
In Fig. 1, the N and F quantities are shown by thick
continuous and dashed curves, respectively.

In a systematic notation explained in Sect. \ref{Sec:3}, 
the results obtained from the Fuller NF method, which substitutes 
(\ref{EulDec}) into (\ref{ParDev}), are indicated by $\text{R} = 0$. 
The thin curve, in the lower panel of Fig. 1, shows the full 
cross section.
Figure 1 shows that, for $\text{R} = 0$, unphysical 
contributions dominate the F (= N) cross section over most 
of the angular range, i.e. for $\theta \gtrsim 50^\circ$.
In particular, the F curve is completely different
from the expected exponential decrease.
Also the F (= $-$N) LAM $\approx 0$ for 
$\theta \gtrsim 50^\circ$. 
At forward angles, oscillations in the F LAM curve 
indicate that another F subamplitude is present, which interferes 
with the unphysical one. 
This behavior does not support the conjecture that the F LAM of 
this other subamplitude has the expected value, $\Lambda = 10.0$.

By comparing the lower panel of Fig. 1 with the corresponding
Fig. 1 of \cite{ANN01d}, for which $\Lambda = 10$ and $\Delta=2.0$, 
one observes that the change in $\Delta$ has not 
altered the magnitude of the unphysical contribution in the angular 
region where it is dominant.
The explanation for this observation is straightforward.
The amplitude obtained on dropping $S_l$ from the term 
$S_l -1$ in (\ref{ParAmp}) is 
$f_\delta(\theta) = i \delta(1-{x})/k$.
Applying to $f_\delta(\theta)$ the same procedure used by 
Fuller to derive the NF Coulomb subamplitudes 
(Ref. \cite{FUL75}, Eq. 11), we find that the NF components 
of $f_\delta(\theta)$ are
$f^{(\mp)}_\delta(\theta) = \pm [2 \pi k (1-{x})]^{-1}$.
The corresponding NF cross sections, downward shifted by 
one vertical unit, are shown in Fig. 1 by the thin dotted curve
and labelled NF$_\delta$.
This curve shows that the major part of the unphysical contribution, 
when it dominates the Fuller F subamplitude, is due to 
$f^{(+)}_\delta(\theta)$
and does not depend on the properties 
of $S_l$.
\begin{figure}
\centering
\includegraphics[width = 8.6 cm]{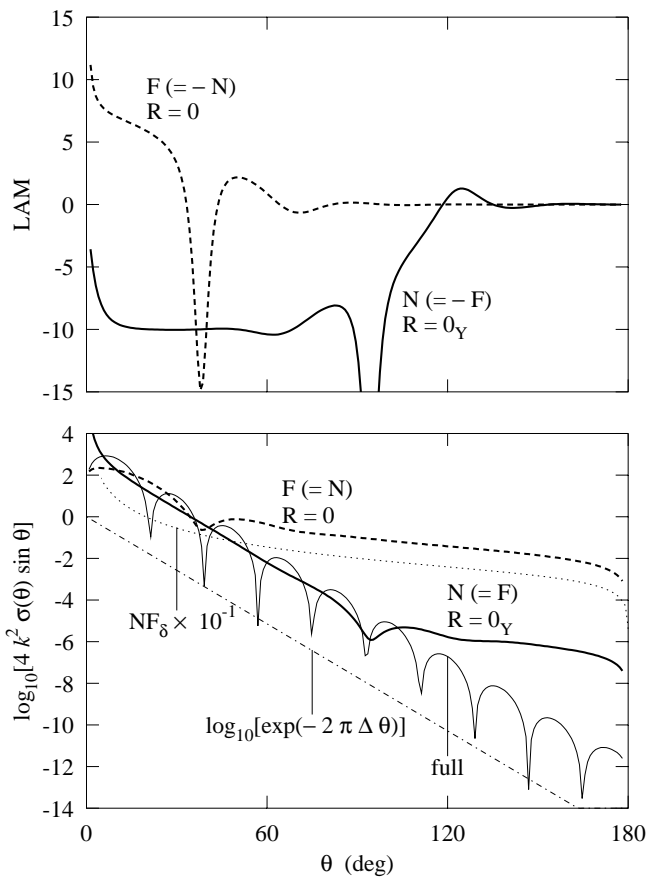}
\caption{\label{fig:Fig01} 
Strong absorption model N (thick continuous curves) and F
(thick dashed curves) cross sections (lower panel)
and LAM (upper panel) calculated using the 
$\text{R} = 0$ and $\text{R} = 0_{\text Y}$
NF methods.
The thin curve shows the full cross section. 
The thin dotted curve (NF$_\delta$) shows the F (= N) 
cross section for the unphysical 
amplitudes $f^{(\mp)}_\delta(\theta)$ (displaced downward by one unit). 
The thin dot-dashed line shows the slope of the expected 
exponential behavior for the NF cross sections}
\end{figure}

A more satisfactory Fuller NF result is obtained
by dropping $1$ from the term $S_l - 1$ in (\ref{ParAmp}),
and then considering the PWS thus obtained as a distribution.
To ensure convergence of the resulting NF subamplitudes, a YRW resummation
is performed on them. 
This is explained in more detail in Sect. \ref{Sec:3},
and in a systematic notation developed there, is denoted 
$\text{R} = 0_{\text Y}$. 
Figure 1 shows that the $\text{R} = 0_{\text Y}$ results are rather 
good at forward angles, $\theta \lesssim 70^\circ$. 
Apart from a small region around $\theta = 0^\circ$,
where the condition $\Lambda \sin \theta \gg 1$ is not 
satisfied, the N (= $-$F) LAM agrees 
closely  with the expected value of $-\Lambda$ up to 
$\theta \approx 70^\circ$, and the N (= F) cross section 
curve follows the expected exponential decrease. 
For $\theta \gtrsim 120^\circ$, the N cross section is still dominated
by an unphysical contribution.
At intermediate angles, $70^\circ \lesssim \theta \lesssim 120^\circ$, 
interference oscillations appear both in the N cross section and in 
the N LAM curve.
It is interesting to note that the NF LAMs are more 
sensitive to interference effects in Fig. 1 than are the 
NF cross sections. 
Also, in the interference region, one cannot attach the meaning
of a {\it local angular momentum} to the subamplitude
phase derivative.
In our case, in this interference region, the N LAM curve 
oscillates around the expected semiclassical value of 
$-\Lambda$ in the region, 
$70^\circ \lesssim \theta \lesssim 90^\circ$, 
where the true semiclassical component dominates the 
N subamplitude, and around the unphysical value of 0 at larger angles.

%------------------
\subsection{Optical model for $^{16}$O $+$ $^{16}$O elastic
scattering}
\label{Sec:22}

Figure 2  shows our results for the phenomenological (WS2) optical potential 
used to fit \cite{KHO00} the $^{16}$O + $^{16}$O elastic 
cross section at $E_{\text {lab}} = 145$ MeV.
The usual Fuller NF method has been applied that employs an
analytic formula for the NF subamplitudes of the Coulomb
scattering amplitude \cite{FUL75}.
The parameters for this potential are given in Table 1
of Ref. \cite{KHO00}.
\begin{figure}
\centering
\includegraphics[width = 8.6 cm]{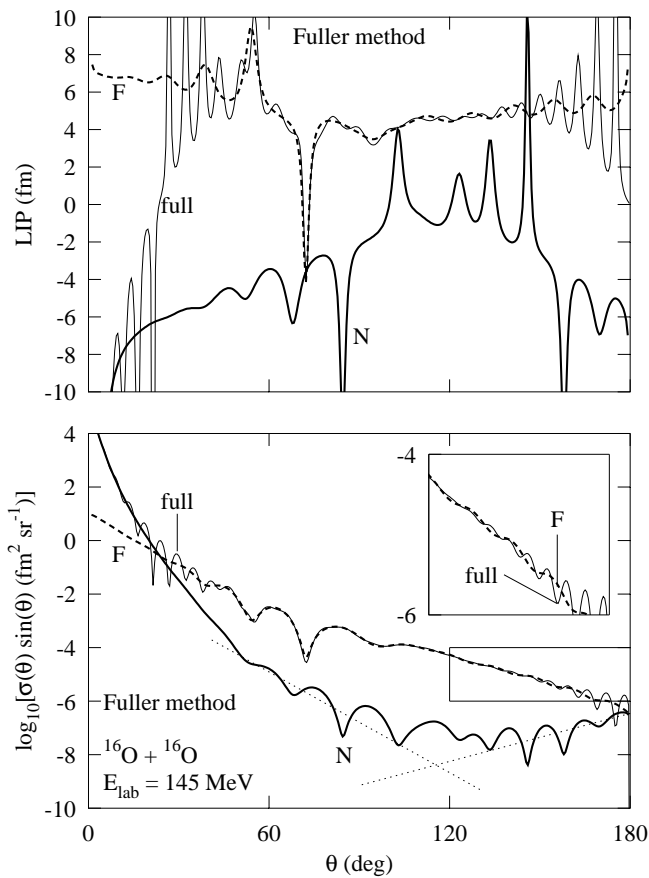}
\caption{\label{fig:Fig02} 
Fuller N (thick continuous curves) and F
(thick dashed curves) cross sections (lower panel)
and LIP (upper panel) for $^{16}$O + $^{16}$O elastic 
scattering at $E_{\text {lab}} = 145$ MeV.
The thin continuous curves show the cross section and LIP 
using the full amplitude.
The two thin dotted lines show interpolations 
of the average behavior of the N cross section 
in the ranges $50^\circ \le \theta \le 90^\circ$
and $\theta \ge 150^\circ$. The inset in the lower panel
show a vertical magnification by a factor of 3 of the thin 
rectangular area.
}
\end{figure}
In the upper panel we display LAM/$k$, which we call 
the {\it Local Impact Parameter} (LIP), and in the lower 
panel a log plot of $\sigma (\theta) \sin \theta$.
The thin continuous lines show the  
(unsymmetrized) cross section and LIP for the full amplitude. 
For $\theta \approx 20^\circ$, and at backward angles,
the full cross section is characterized by oscillations
having approximately the same period.
In the intermediate angular range, $50^\circ \lesssim \theta
\lesssim 150^\circ$, there are oscillations of longer period,
which increases as $\theta$ increases.
These properties, which are typical of light heavy-ion 
scattering, are much more complicated than those for the 
simple SAM cross sections, and indicate the presence of several 
interfering subamplitudes.

The oscillations in the LIP  curve for the full amplitude
qualitatively follow those in the full cross section.
In addition, the change in sign of the full LIP for
$\theta \approx 20^\circ$ indicates a crossover between one N 
subamplitude, dominant at smaller angles (where the full LIP oscillates 
around negative values), with a F subamplitude that is dominant 
at larger angles
(where the full LIP oscillates around positive values). 
The oscillations of longer period in the intermediate 
angular range for the full cross section correspond 
to an oscillating positive LIP.
This indicates there is an interference between two F 
subamplitudes with a crossover near the  
deep minimum at $\theta \approx 70^\circ$.
The oscillations at backward angles in the full cross section 
indicate an interference of 
one of these F subamplitudes with its continuation to 
the opposite side of the target (N).

The results of the Fuller NF method (thick curves,
continuous for N and dashed for F quantities) in Fig. 2
confirm this interpretation.
The N cross section crosses the F one at $\theta \approx 20^\circ$;
the F cross section oscillates in the intermediate 
angular range before decreasing, almost 
monotonically, towards backward 
angles, where it meets the N cross section. 
For $\theta \gtrsim 60^\circ$, the behavior of the N 
cross section is characterized by rather complicated oscillations.
It is difficult to imagine that these oscillations
arise only from interference between a N subamplitude, 
which is dominant at forward angles, with a N 
subamplitude dominant at backward angles. 
The thin dotted lines in Fig. 2 show interpolations 
(linear fits) of the average behavior of the N cross 
section in the ranges $50^\circ \le \theta \le 90^\circ$
and $\theta \ge 150^\circ$.
In the crossing region of the dotted lines, the N cross
section is about two orders of magnitude larger then 
the crossing value.
The contribution from an additional N subamplitude is 
apparently necessary to explain this behavior of the N cross 
section.
The shape of the N LIP curve supports this conjecture.
The N LIP curve is oscillatory, with 
increasing oscillation amplitude, up to $\theta \approx 85^\circ$, 
where the deep minimum corresponds to a crossover  
between two interfering 
N subamplitudes.
For $85^\circ \lesssim \theta \lesssim 145^\circ$,
the N LIP curve is oscillatory
around 0, which suggests an unphysical origin for the N subamplitude
dominant in this angular range.
At $\theta \approx 145^\circ$, where there is a narrow maximum 
in the N LIP curve, the unphysical N subamplitude crosses a different 
N subamplitude, that becomes dominant at backward 
angles.
Oscillations, similar to those observed at backward angles
in the N LIP curve, are present at 
backward angles in the F LIP curve.
The corresponding oscillations in the F cross section are 
barely visible with the scale used in the lower panel of Fig. 2.
They can, however, be observed in the inset, 
where the thin rectangular area is plotted with a 
vertical scale magnified by a factor of 3.

In summary our detailed analysis of the NF LIP and NF cross 
sections indicates the
presence of an unphysical contribution, which makes the 
properties of these quantities unneccessarily complicated.

%------------------
\subsection{Additional optical model examples}
\label{Sec:23}

\begin{figure}
\centering
\includegraphics[width = 8.6 cm]{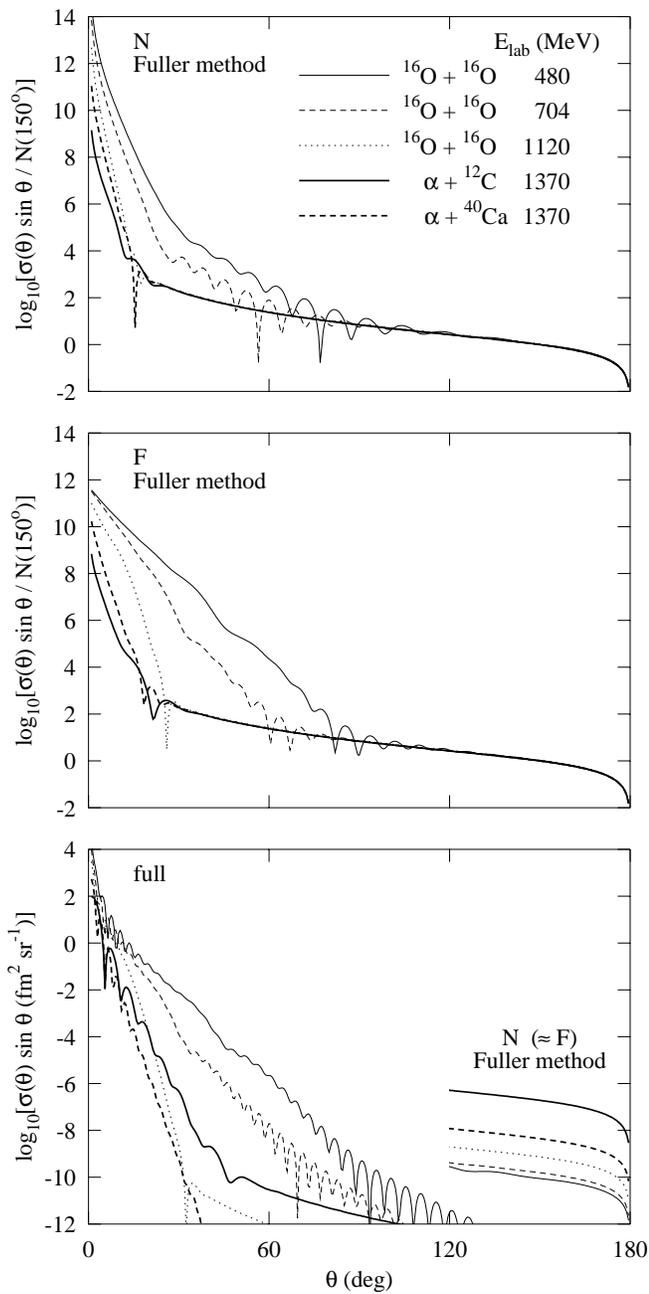}% Here is how to import EPS art
\caption{\label{fig:Fig03} 
Fuller N (upper panel) and F
(middle panel) optical potential cross sections for $^{16}$O + $^{16}$O, 
$\alpha$ + $^{12}$C and $\alpha$ + $^{40}$Ca at different
energies.
All the N and F curves are normalized to the values of the 
corresponding N quantities at $\theta = 150^\circ$.
The lower panel shows the full cross sections and, for
$\theta > 120^\circ$, the unphysical behavior of the
Fuller N ($\approx$ F) cross sections.}
\end{figure}

In this section we briefly discuss five additional
examples where the NF optical potential cross sections
are dominated by unphysical
contributions over wide angular ranges.

Figure 3 shows N (upper panel) and F (middle
panel) cross sections obtained by the usual Fuller 
NF method for different optical potentials
that describe the elastic scattering of $^{16}$O
by  $^{16}$O, at $E_{\text {lab}}$ = 480, 704 and 1120
MeV (potentials labelled WS2 in Ref. \cite{KHO00}), and of $\alpha$ 
particles by $^{12}$C and $^{40}$Ca, at  
$E_{\text {lab}}$ = 1370 MeV (potentials labelled 
WSa in Ref. \cite{KHO02}).
We used relativistically corrected
kinematics \cite{FAR84} in all these optical potential
calculations, and symmetrisation effects were ignored
for the $^{16}$O + $^{16}$O system.

In the N and F panels, we have normalized the
quantities, $\log_{10}[\sigma(\theta) \sin \theta]$, 
to the values of the corresponding
N quantities, indicated by $N(\theta_0)$, at $\theta_0=150^\circ$.
Except for a restricted region at forward
angles, whose width depends on the system considered, 
the normalized Fuller NF cross sections are practically equal, 
for different colliding partners and/or energies.
In every case, the unphysical effects of the
Fuller NF method are as striking as those 
for the SAM discussed in Sect.
\ref{Sec:21}.
In the lower panel we have plotted,
without normalization factors, the full 
cross sections and, for $\theta \ge 120^\circ$,
the Fuller N ($\approx$ F) cross section.
It is very apparent that, at large angles, the 
behavior of the full cross sections is quite different 
from that of the Fuller NF cross sections.
In particular, the $^{16}$O + $^{16}$O full cross
sections at 480 MeV and 704 MeV possess a pronounced interference 
pattern in an angular region where the Fuller NF cross sections
are dominated by unphysical contributions.
For completeness, we remark that the NF LIP is 
practically null in the angular ranges where the NF
cross sections show unphysical behavior for all 
these cases. 

%------------------
\section{Improved NF method using resummation}
\label{Sec:3}

%------------------
\subsection{Resummation of full and NF PWS} 
\label{Sec:30a}

In the preceding Section, we have discussed examples where the
Fuller NF method resulted in unphysical contributions.
We also clearly demonstrated that
the Fuller method has the capability to identify the unphysical 
contributions through anomalous behavior in the NF LAM, or LIP. 
This capability helps us to avoid misleading interpretations 
of the full and NF cross sections obtained from the Fuller NF method. 
However the problem of finding more satisfactory NF 
methods without unphysical contributions remains open.

A possible solution to this problem was proposed by Hatchell 
\cite{HAT89}, who used a modified NF method.
The modifications consisted of, first, in writing $f(\theta)$ 
in the resummed form ($x \ne 1$) 
\begin{equation}
\label{JenDev}
f(\theta )=\frac{1}{2 i k} \frac{1}{(1-{x} )^{r}}
\sum_{l=0}^{\infty} a_{l}^{(r)}P_{l}({x} ),
\end{equation}
$r =1,2,\ldots$, and, second, in using a different splitting
in the resummed PWS (\ref{JenDev}) for the Legendre polynomials into traveling 
waves. 

The use of the resummed form (\ref{JenDev}) for $f(\theta)$ was 
originally proposed \cite{YEN54} by Yennie, Ravenhall, and Wilson 
(YRW) in a different context.
Equation (\ref{JenDev}) is an exact resummation formula, of order $r$,
which is derived from the recurrence relation for Legendre polynomials.
Some mathematical properties of the resummed PWS
(\ref{JenDev}) have been investigated by Wimp \cite{WIM99}.
The YRW resummed form (\ref{JenDev}) can be derived by iterating 
$r$ times, starting from $a_{l}^{(0)}=a_{l}$, the 
resummation identity ($x \ne 1$)
\begin{equation}
\label{JenIde}
\sum_{l=0}^{\infty} a_{l}^{(i-1)}P_{l}({x})
=
\frac{1}{1 - {x}}
\sum_{l=0}^{\infty} a_{l}^{(i)}P_{l}({x}),
\quad i=1,2,\ldots,
\end{equation}
where
\begin{equation}
\label{JenRec}
a_{l}^{(i)}=-\frac{l}{2l-1} a_{l-1}^{(i-1)}+a_{l}^{(i-1)}-\frac{l+1}{2l+3}%
a_{l+1}^{(i-1)},
\end{equation}
with $a_{-1}^{(i-1)}= 0$. 

It is important to remark that the resummed coefficients $a_l^{(r)}$, 
being linear combinations of the $a_l^{(0)} = a_l$, 
can have very different properties from the original physical 
$a_l$.
However, no information about the physical 
$a_l$ is lost on applying the resummation procedure, and the 
value of $f(\theta)$ does not depend in any way on the
resummation order used.
This is true for (\ref{JenDev}) and for all the resummed
forms for $f(\theta)$ 
derived in this paper and is a consequence of the 
(exact) mathematical properties of
the Legendre polymonials.

Equation (\ref{JenDev}) does not hold for a PWS written in terms 
of a linear combination of Legendre 
functions of integer degree of the first and second kinds ($x \ne \pm 1$) 
\begin{equation}
\label{PWSe}
{\cal F}(\theta) = \sum_{l=0}^{\infty} a_l {\cal L}_l (x),
\end{equation}
where 
\begin{equation}
{\cal L}_l (x) = p P_l (x) + q Q_l(x), 
\end{equation}
with $p$ and $q$
real or complex constants independent of $l$.
Rather, as a result of the property $l Q_{l-1} ({x}) \rightarrow 1$ as 
$l \rightarrow 0$, the {\it extended} resummation identity holds
($x \ne \pm 1$) \cite{ANN81}
\begin{equation}
\label{JenIdeE}
\sum_{l=0}^{\infty} a_{l}^{(i-1)}{\cal L}_{l}(x)
=
\frac{1}{1 - x} \left [
\sum_{l=0}^{\infty} a_{l}^{(i)}{\cal L}_{l}(x)
- q a_0^{(i-1)} \right ],
\end{equation}
where the recurrence relation (\ref{JenRec}) gives 
$a_l^{(i)}$ in terms of $a_l^{(i-1)}$.
By iterating (\ref{JenIdeE}) $r$ times ,  ${\cal F} (\theta)$ 
can be written in the {\it extended} resummed form 
($x \ne \pm 1$) \cite{ANN81}
\begin{equation}
\label{JenDevE}
{\cal F}(\theta )=\frac{1}{(1- x)^r} 
\sum_{l=0}^{\infty} a_{l}^{(r)} {\cal L}_{l}(x)
- q \sum_{i=1}^r a_0^{(i-1)} \frac{1}{(1- x)^i}.
\end{equation}
It is important to note that ${\cal F}(\theta)$ is independent of
the value used for $r$.

The $Q_l^{(\mp)} ({x})$ used to split
$P_l ({x})$ in the Fuller NF method are a 
particular case (with $p= 1/2$, and $q = \pm  i / \pi$) 
of the more general linear combination  ${\cal L}_l ({x})$

When the splitting (\ref{EulDec}) is made in the initial (\ref{ParDev})
or resummed PWS (\ref{JenDev}), different NF subamplitudes are
obtained, i.e. the NF subamplitudes depend on the order
$r$ of the resummation.
In particular, the last term on the r.h.s. of (\ref{JenDevE}) is 
omitted from these $r$ dependent resummed NF subamplitudes.
However, the sum of the NF resummed subamplitudes remains $f(\theta)$,
because the differences exactly cancel each other.
A different extended resummation identity
occurs in the Hatchell approach, because the functions he 
used in place of the $Q_l({x})$ obey a (inhomogeneous) 
recurrence relation, different from that for $Q_l({x})$ 
(see Ref. \cite{HOL99a} for details).

It is important to realize that the YRW resummation
procedure (which has been extended in \cite{ANN81} to treat 
PWS like (\ref{PWSe}) )
can also be used to speed up the convergence of the PWS
(\ref{ParDev}) or (\ref{ParDevmp}),
or even to ensure their convergence if the PWS was originally defined
only in a distributional sense: in fact if $a_l \sim O(l^{-p})$
then $a_l^{(r)} \sim O(l^{-p-2r})$.
Indeed the YRW resummation procedure was originally introduced
\cite{YEN54} 
to produce, for  $r \ge 1$, a convergent \cite{GOO80} 
PWS for the full amplitude of high-energy electron-nucleus 
scattering in which $S_l$ is asymptotically Coulombic,
and also to speed up the convergence of this PWS.

Using his splitting for $P_l(x)$ into traveling angular waves, Hatchell
has shown \cite{HAT89} that the unphysical contributions to 
the SAM NF cross sections systematically decrease on increasing 
$r$, i.e. on using the resummed PWS (\ref{JenDev}) before the
splitting of $P_l(x)$, rather then the original unresummed
PWS (\ref{ParDev}).
% if one perform the splitting of $P_l({x})$ in running 
%angular waves in a $r$ order resummed PWS (\ref{JenDev}) rather 
%than in the original one (\ref{ParAmp}).
%
More recently \cite{HOL99a,HOL99b}, it was shown that the
same NF resummed method gives even better results
if the Fuller $Q_{l}^{(\mp)}({x})$ functions are used.
The superiority of the $Q_{l}^{(\mp)}({x})$  
seems to be connected with the greater rapidity with which the 
$Q_{l}^{(\mp)}({x} )$ approach their asymptotic behavior 
(\ref{AsyQpm}) \cite{McC95,HOL99a,HOL99b}, compared to
the Hatchell NF functions.

The success of using (\ref{JenDev}) before
applying the NF splitting (\ref{EulDec}) depends upon
the properties of the resummed coefficients $a_l^{(r)}$.
For the SAM at low $l$ values, the $a_l^{(r)}$ 
rapidly decrease in magnitude \cite{HOL99a} with increasing $r$.
As a result, the low $l$ values, where the 
splitting into running angular waves is physically 
less reasonable, give a smaller contribution to the
resummed PWS.

However, in some cases, this resummation technique acts in the 
opposite direction, by enhancing the undesired unphysical 
contributions to the NF resummed subamplitudes.
We have found that this happens, for example, 
for pure Coulomb scattering, for scattering by an impenetrable 
sphere, and for the SAM (see \cite{NOL02}) when the cross section 
is calculated at an angle $\pi -\theta$, using the identity 
$P_l[\cos (\pi-\theta)]=(-1)^l P_l(\cos \theta)$.

%------------------
\subsection{Improved resummation of full and NF PWS} 
\label{Sec:30b}

One possible solution to the intriguing puzzle discussed
in Section \ref{Sec:30a} is to regard 
(\ref{JenIde}) as a particular case of the modified
resummation identity \cite{WHI01}
\begin{equation}
\label{JenIdeGen}
\sum_{l=0}^{\infty} a_{l}^{(i-1)}P_{l}({x})
=
\frac{1}{\alpha_{i} + \beta_{i} {x}}
\sum_{l=0}^{\infty} a_{l}^{(i)}P_{l}({x}),
\quad i=1,2,\ldots,
\end{equation}
with $\alpha_i+ \beta_i {x} \ne 0$ and
\begin{equation}
\label{JenRecGen}
a_{l}^{(i)}= \beta_{i} \frac{l}{2 l-1} a_{l-1}^{(i-1)}
+ \alpha_{i} a_{l}^{(i-1)}
+ \beta_{i} \frac{l+1}{2 l+3} a_{l+1}^{(i-1)}.
\end{equation}
For $\alpha_i, \beta_i \ne 0$, apart a renormalization factor, the  
r.h.s. of
(\ref{JenIdeGen}) depends only on the ratio $\beta_{i}/\alpha_{i}$.
Thus, without loss of generality, we can assume $\alpha_{i}=1$
for all $i$.
By iterating (\ref{JenIdeGen}) $r$ times, we can write $f(\theta)$
in the modified resummed form ($1+\beta_i \ne 0$)
\begin{equation}
\label{JenDevGen}
f(\theta )=\frac{1}{2 i k}
\left ( 
\prod_{i=1}^{r} \frac{1}{1+\beta_i {x} }
\right )
\sum_{l=0}^{\infty} a_{l}^{(r)}P_{l}({x} ),
\end{equation}
$r = 1, 2, \ldots \, $. It is straightforward to show that the extended 
modified resummed form for the PWS (\ref{PWSe}) is given by
($x \ne \pm 1$ and $ 1+\beta_i \ne 0$)
\begin{eqnarray}
{\cal F}(\theta ) &=&
\left ( 
\prod_{i=1}^{r} \frac{1}{1+\beta_i {x} }
\right )
\sum_{l=0}^{\infty} a_{l}^{(r)}{\cal L}_{l}({x} ) \nonumber \\
&+&q\sum_{i=1}^r \beta_i a_0^{(i-1)} 
\prod_{j=1}^i \frac {1}{1+\beta_j x}.\label{JenDevExt}
\end{eqnarray}
The identity (\ref{JenDevGen}) is the key result for our
improved NF method.
The YRW resummation formula (\ref{JenDev}) is  
obtained when $\beta_{1} = \beta_{2}= \ldots = \beta_{r} = -1$.

The resummed form (\ref{JenDevGen}) with 
$\beta \equiv \beta_1 =\ldots=\beta_r$ is a particular case of a
more general one \cite{WHI01}, which uses a basis set
of reduced rotation matrix elements; this gives 
the amplitude for more general scattering processes
than those described by (\ref{ParDev}).
For these general PWS, a Fuller-like NF decomposition
can be introduced \cite{ABI99,SOK99,McC01} which
allows the scattering amplitude to be separated into 
NF subamplitudes. 
In some cases, the NF cross sections contained 
unexpected (unphysical) oscillations \cite{WHI01}, 
which are enhanced if the generalization of 
(\ref{JenDev}) is used, but which disappear for 
an appropriate choice of the $\beta$-parameter in 
the generalization of (\ref{JenDevGen}).

The considerable successes achieved by the original Fuller 
NF method suggest that the modified
resummed form (\ref{JenDevGen}) be used to diminuish 
unphysical contaminations to the NF subamplitudes
when they are present.
To do this, we must give a practical rule to fix the values
of the $\beta_i$ parameters.
In Refs. \cite{NOL02,WHI01}, it was proposed to select the 
value of $\beta \equiv \beta_1 =\ldots=\beta_r$ so that 
$(1+\beta {x})^{-r}$  {\it approximately mimics 
the shape of the angular distribution}.
The shape of the cross section can however be very
different from that given by 
$(1+\beta {x})^{-r}$.
It is therefore desirable to test a different recipe,
based on a simple rule.
The quantitative recipe proposed here is inspired by the 
observation that the modified resummation procedures produce 
a more physical NF understanding by reducing the contribution 
from the low $l$ values in the resummed PWS.
This suggests that we select the $\beta_1, \beta_2, \ldots, 
\beta_r$ in $r$ repeated applications of (\ref{JenIdeGen}), 
so as to eliminate as many low $l$ terms as possible 
from the resummed PWS in  (\ref{JenDevGen}).
The transformation from $\{a_l^{(i-1)}\}$ to $\{a_l^{(i)}\}$
is linear tridiagonal, with coefficients linear in
$\beta_{i}$, which means  that a resummation of order $r$  
allows one to equate to zero the leading $r$ 
coefficients $a_l^{(r)}$, with $l=0,1,\ldots,r-1$, by solving 
a system of $r$ equations of degree $r$ in
the parameters $\beta_1, \beta_2, \ldots, \beta_{r}$.
We will call the resummation defined in this way an
{\it improved resummation of order $r$}.

It is straightforward to show that the improved 
resummation of order $r = 1$ is obtained by choosing
\begin{equation}
\beta_1=-3 a_0 / a_1,
\label{OneOpt}
\end{equation}
while the improved resummation of order $r=2$ is given by
\begin{equation}
\beta_{1,2}=(B \pm \sqrt{B^2-4A}) /2 \, ,
\label{TwoOpt}
\end{equation}
with $A$ and $B$ solutions of the linear equations
\begin{equation}
\left \{
\begin{array}{lcr}
(\frac{1}{3} a_0+\frac{2}{15} a_2) A &+  &\frac{1}{3} a_1 B =-a_0\,
\\
\\
(\frac{3}{5} a_1+\frac{6}{35} a_3) A &+ &(a_0+\frac{2}{5} a_2) B =-a_1.
\end{array}
\right.
\label{SYS}
\end{equation}
Higher order improved resummations require the solution of more
complicated systems of equations.

Note that the improved resummation of order $r = 1$ is, obviously, not
defined if $a_1=0$. Similarly the resummation of order $r=2$ 
is not defined for an (accidentally) zero value of the 
determinant of the linear equations (\ref{SYS}). 
Analogous limitations 
are expected to hold for higher order resummations.
However, these pathological situations were never observed in
our calculations.
In all the cases we have analyzed using $r \leq 2$, we find that the
improved resummations considerably reduce the width of the 
angular regions in which the Fuller NF cross sections exhibit 
unphysical behavior.
In these analyses, we have used $S_l$ from simple parametrizations,
as well as from some of the optical potentials currently employed 
to describe light heavy-ion scattering.

Note that, as for the YRW resummed form (\ref{JenDev}), no physical
information is lost on using our improved resummed
form (\ref{JenDevGen}), and the full $f(\theta)$ does not depend in any way
on the resummation order used.
This might seem surprising, because our method 
omits the contribution from low values of $l$.
For example, for the improved resummation of order 
$r=1$, the {\it resummed} term $a_0^{(1)}$ is set to zero in the {\it resummed}
PWS, but the information on the {\it physical} value of $a_0$ is contained 
in the {\it resummed} term $a_1^{(1)}$ and in the
resummation parameter $\beta_1$, and similarily
for higher order resummations.
The improved resummation method only modifies the 
NF (resummed) subamplitudes, not their sum $f(\theta)$, 
as it tries to eliminate unphysical contributions
that have their origin in the NF splitting
of the $P_l(x)$ into running
waves $Q_l^{(\mp)}(x)$.

The physical meaning of the NF splitting (\ref{EulDec}) is based on 
the asymptotic properties of $P_l({x})$ and 
$Q_l^{(\mp)}({x})$, see (\ref{AsyQpm}).
The NF splitting (\ref{EulDec}) is not expected to
be physically meaningful for low $l$ values.
For example, for $l = 0$, (\ref{EulDec}) states that
$P_0({x}) = 1$ is the sum of
\begin{equation} 
Q_0^{(\mp)}({x}) = \frac{1}{2} \pm \frac{i}{2 \pi}
\ln[(1+x)/(1-x)].
\end{equation}
This decomposition of unity is undoubtly mathematically exact, but
it is difficult to think that it can be physically
meaningful.
It is contributions of this type that the improved 
resummation omits from the resummed PWS, after
having moved to higher values of $l$ in the
resummed PWS, the physical 
information contained in the original $S_l$. 

It is also interesting to understand the reasons for the 
successes and failures of the Hatchell
method for the SAM \cite{HOL99a,HOL99b,NOL02} based  on the use of the YRW 
resummed form (\ref{JenDev}) for $f(\theta)$ and the Fuller NF
splitting (\ref{EulDec}).
In those cases where the Hatchell method is successfull,
we find for low $l$ values, $S_l \approx S_{l+1}$, and the $a_l^{(r)}$
are made small in magnitude with the choice 
$\beta_1 \approx \beta_2 \approx \ldots \approx \beta_r \approx -1$.
In those cases where the Hatchell method fails, it happens that,
for low $l$ values, $S_l \approx -S_{l+1}$, and the choice
$\beta_1 \approx \beta_2 \approx \ldots \approx \beta_r \approx 1$
makes the $a_l^{(r)}$ small in magnitude.
Our improved method automatically chooses appropriate values 
for the resummation parameters using the exact values of
$S_l$. 

Finally, we summarize the procedure used by our new improved 
resummation method.
The calculations are performed applying:
{\it first}, an improved resummation of order 
$r = 0, 1, 2$, with $r = 0$ meaning no resummation, 
{\it second}, the Fuller NF splitting (\ref{EulDec}), 
and {\it third} an extended YRW resummation (\ref{JenDevE}) of 
the NF (resummed) subamplitudes.
This latter YRW resummation is necessary to
ensure the convergence of the final NF PWS.
This final resummation can, however, be replaced with any 
other procedure that provides convergence of the final NF PWS.
The results obtained from these three steps will be indicated
by the notation $\text{R} =0_{\text Y}, 1_{\text Y}, 2_{\text Y}$.
The notation $\text{R} =0_{\text Y}$ is used in Fig. 1.
Also, we use $\text{R} = 0$ to indicate the original Fuller method
for the SAM case: no 
resummation and no final YRW resummation, because, 
in this case, the presence of the 1 in the term  $S_l -1$ ensures 
rapid convergence of the PWS.

%---------------------
\subsection{Strong absorption model for elastic scattering}
\label{Sec:31}

Our results obtained after applying the 
$\text{R} = 1_{\text Y}, 2_{\text Y}$ 
procedures  to the
SAM of Sec. \ref{Sec:21} are shown in Fig. 4.
\begin{figure}
\centering
\includegraphics[width = 8.6 cm]{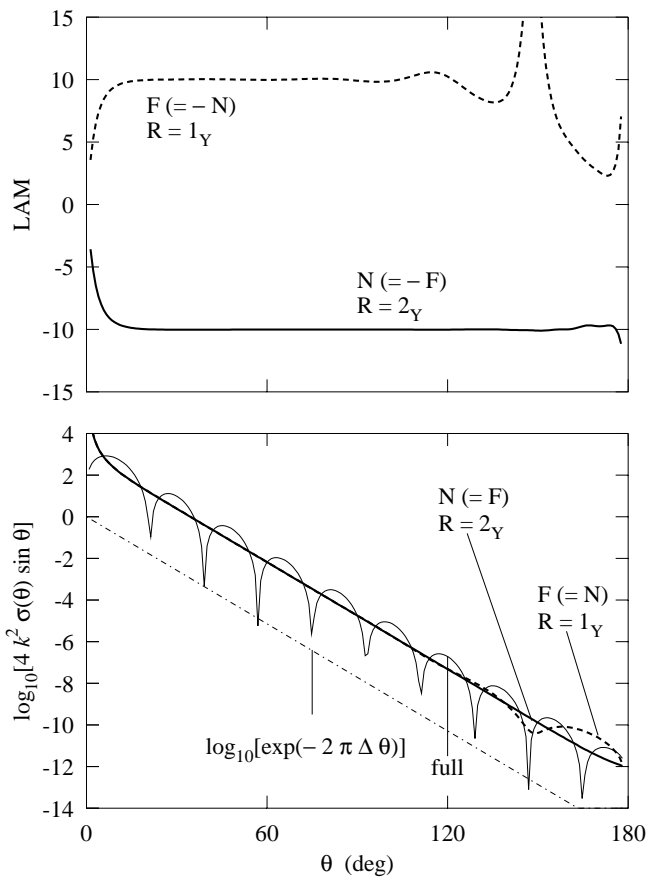}
\caption{\label{fig:Fig04} 
Same as Fig. 1, except using the 
$\text{R} = 1_{\text Y}, 2_{\text Y}$ resummation methods.
}
\end{figure}
The effectiveness of the improved resummation procedure
is impressive, both for the NF cross sections and for the NF LAM
curves. 
Using the $\text{R} =1_{\text Y}$ method 
(for which $\beta_1=-0.7621$), 
%(for which $\beta_1=-0.800$) %for \Delta=2.0
%(for which $\beta_1=-0.831$) %for \Delta=2.2
the F (= $-$N) LAM and the F (= N) cross section are in agreement with 
the expected results up to $\theta \approx 140^\circ$. 
For $\text{R} =2_{\text Y}$ (which has 
%$\beta_{1, 2}=-0.879 \pm 0.076 \, i$), %for \Delta=2.0
%$\beta_{1, 2}=-0.892 \pm 0.076 \, i$), %for \Delta=2.2
$\beta_{1, 2}=-0.8634 \pm 0.0723 \, i$),
the agreement covers almost the whole angular range. 
The small irregular oscillations appearing at large 
$\theta$ for the N LAM curve, when 
$\text{R} =2_{\text Y}$, may arise from precision 
limitations (64 bit floating point representation) in the 
calculations, or from residual unphysical contributions
not completely removed by our improved method.

If instead we apply the YRW resummation 
procedure, we find that it gives worse results 
compared to our $\text{R} =1_{\text Y}, 2_{\text Y}$
methods, although better results than the Fuller 
$\text{R} =0_{\text Y}$ method.
This is because the YRW choice of $\beta_1=\beta_2=
-1$ is not too far from our  improved estimate of these parameters.

Note that, using our improved procedure, the value of the 
SAM cut-off parameter, $\Lambda=10$, is correctly identified 
in the plots of the LAM.
Also, no shift of the LAM occurs on changing the resummation
order.
We recall that resummation does not 
change or translate in any way the physical properties  
$S_l$; nor does it alter the physical content 
of $f(\theta)$.
It only changes the coefficients $\alpha_l^{(r)}$ in the 
resummed PWS (\ref{JenDevGen}). 
Note that the $\alpha_l^{(r)}$ have a different meaning
from the partial wave amplitudes $a_l$, and the summation 
index of the resummed PWS (\ref{JenDevGen}) should 
not be identified as the orbital angular momentum quantum 
number.
This is because the resummation extracts the factor
$\prod_{i=1}^r (1+\beta_i x)^{-1}$ from the Legendre 
polynomial PWS.

%------------------
\subsection{Optical model for $^{16}$O + $^{16}$O
elastic scattering}
\label{Sec:32}

\begin{figure}
\centering
\includegraphics[width = 8.6 cm]{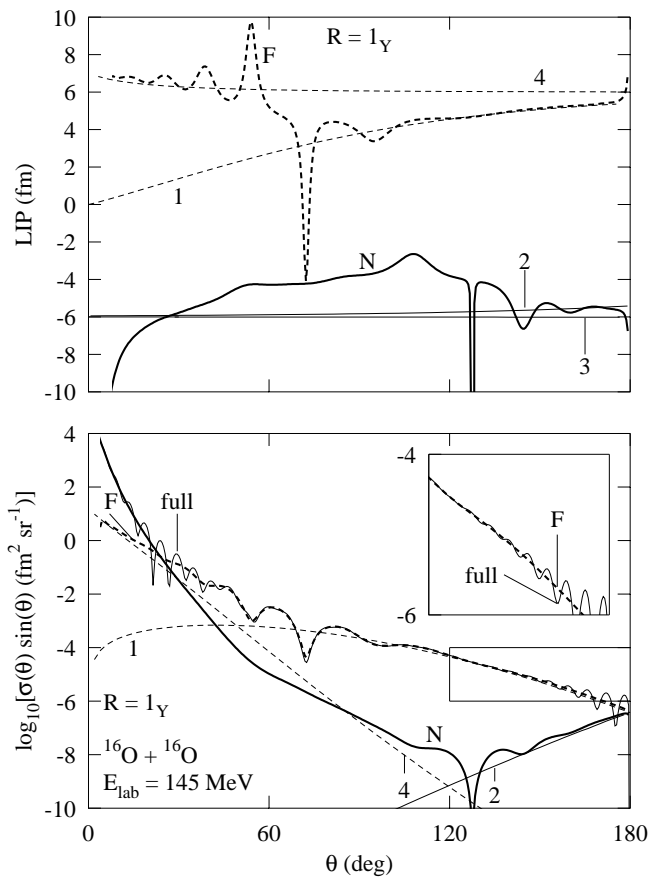}
\caption{\label{fig:Fig05} 
Same as Fig. 2 except using the $\text{R} = 1_{\text Y}$ NF 
resummation method and the full LIP curve is not shown.
Cross sections and LIP calculated using classical mechanics are shown
by thin curves (continuous for N and dashed for F contributions). 
The indices 1, 2, 3 and 4 identify the curves corresponding to different branches of 
the classical deflection function.}
\end{figure}

Figure 5  shows our results for the WS2 optical potential of Sec. \ref{Sec:22} 
using the $\text{R} =1_{\text Y}$ method (for which 
$\beta_1=-0.9997-0.0800 \, i$). 
The oscillations of the Fuller N curves around 
$\theta \simeq 90^\circ$ are 
removed, and both the $\text{R} =1_{\text Y}$ N cross section 
and the N LIP curve are considerably simpler than those 
from the Fuller method.
The F curves are essentially the same as those in Fig. 2,
although the $\text{R} =1_{\text Y}$ method has suppressed
the oscillations in the F cross section at backward angles
(compare the insets in the lower panels of Figs. 2 and 5),
as well as the oscillations in the F LIP curve in the same 
angular range.
This demonstrates that the $\text{R} =1_{\text Y}$ method 
is more effective than the usual Fuller one in decomposing 
$f(\theta)$ into slowly varying NF subamplitudes.
It also shows that some oscillations in the Fuller NF cross 
sections in Fig. 2 are artifacts, without any physical
meaning, introduced by the properties of the NF technique used.
We have also applied the improved resummation method 
$\text{R} =2_{\text Y}$. 
The results are practically the same as those for 
$\text{R} =1_{\text Y}$ and are not shown.

The cleaning by the $\text{R} =1_{\text Y}$ procedure 
in Fig. 5 of the original Fuller NF subamplitudes 
(Fig. 2) is impressive and allows 
a better identification, in the F cross section and F LIP
of the dominance, for $\theta \gtrsim 90^\circ$, of the contributions
from classical-like trajectories refracted from the internal
part of the nuclear interaction.
In Fig. 5,  this interpretation is demonstrated 
by the overall agreement, for $\theta \gtrsim 90^\circ$, between the F 
curves and the classical mechanical results (thin curve, labelled 1) 
corresponding to impact parameters smaller than the
classical orbiting one.

For the WS2 potential the collision energy is, in fact, slightly below the
critical energy at which orbiting disappears for the classical 
deflection function (when it is transformed into a nuclear rainbow 
minimum).
Because of this, the dependence of the impact parameter $b$ on $\theta$ 
is that of an infinitly many-valued function (\cite{NEW82}, p. 127-129).
Four of these branches, corresponding to the deflection
function being larger than $-360^\circ$, are plotted for the LIP 
in the upper panel of Fig. 5 (shown by thin curves) and labelled 
from 1 to 4 for increasing values of $b$.
The LIP is assumed to be equal to $b$ for the F branches 
(thin dashed curves, with labels 1 and 4) and to $-b$ in
the N case (thin continuous curves, 2 and 3).
In the lower panel, the thin curves, with the same labels, show 
the contributions to the classical cross section from these 
branches, in which we have included, in the 
usual simple way (\cite{BRI85}, p. 49), the absorptive effects 
of the imaginary part of the optical potential.

The agreement between the quantum $\text{R} =1_{\text Y}$ F 
curves and the classical curve labelled 1 is overall good
for  $\theta \gtrsim 90^\circ$, and is impressive for $\theta \gtrsim 120^\circ$.
This shows that the $\text{R} =1_{\text Y}$ F subamplitude, 
calculated by an exact quantum method, is dominated for 
$\theta \gtrsim 90^\circ$ by a classical contribution 
corresponding to trajectories with small impact parameters,  
$b \lesssim 5$ fm, which are refracted 
by the nuclear interaction.
The contribution  from F branch 1 continues at $180^\circ$ 
into the contribution from N branch 2.
However, as $\theta$ decreases from $180^\circ$, there is an
increasing disagreement with the average quantum N contribution.
Evidently, for impact parameters that approach the orbiting one 
($b_{\text {orb}} = 6.001$ fm), diffractive effects start
to become significant. 
Finally, we note that there is a large disagreement between
the classical F curve 4 and the average behavior of the quantum F cross 
section at forward angles.
This suggests that the other F subamplitude responsible for
the interference pattern in the F quantum cross section cannot be considered
a classical-like refractive contribution.
%   

%------------------
\subsection{Additional optical model examples}
\label{Sec:33}

\begin{figure}
\centering
\includegraphics[width = 8.6 cm]{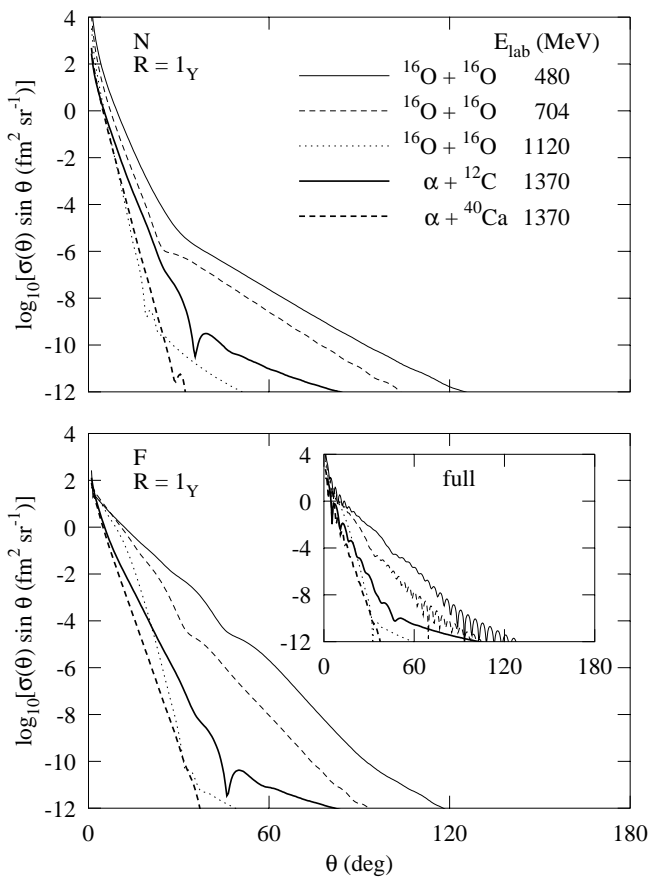}% Here is how to import EPS art
\caption{\label{fig:Fig06} 
Same as for the upper and middle panels of Fig. 3, 
except without normalization factors and the 
$\text{R} = 1_{\text Y}$ NF resummation method has been used.
The inset shows the full cross sections.}
\end{figure}

In Fig. 6, we show the N (upper panel) and F (lower
panel) cross sections calculated using the improved 
NF method, ${\text R} = 1_{\text Y}$, for the 
optical potentials considered in Sec. \ref{Sec:23}.
In contrast to Fig. 3, no normalization factors are 
used in Fig. 6.
Comparison of Fig. 6 with Fig. 3 shows that the 
${\text R} = 1_{\text Y}$ NF 
resummation method has removed
the unphysical contributions that dominate the Fuller 
NF cross sections over most of the angular range.

For these cases, $\beta_1 = -1.0001 - 0.0283 i, -1.0001 - 0.0218 i,
-0.9991 - 0.0102 i, -0.9958 - 0.0021 i$, and $-0.9985 - 0.0010 i$, in 
descending order for the collision systems listed
in the legend of the upper panel of 
Fig. 6.
In all cases, $\beta_1 \approx -1$, which is a 
consequence of the fact that, for these five
collision systems, at low $l$ values we have 
$S_l \approx S_{l+1}$.
The result $\beta_1 \approx -1$ also explains why
the unphysical contributions in the Fuller NF cross 
sections of Fig. 3 have very similar shapes.
Indeed for $\beta_1=-1$, the angular dependence of 
the unphysical cross sections is  
$\propto (1-x)^{-2} \propto \sin^{-4} (\theta / 2)$,
which is the same as the Rutherford
cross section.
This implies that on making the (usual) plot of
the ratio of the cross sections to the Rutherford one, the
unphysical contributions will appear independent of 
$\theta$, i.e. a constant.

Figure 6 clearly shows that the ${\text R} = 1_{\text Y}$
method has cleaned the Fuller NF cross 
sections of unphysical contributions.
The ${\text R} = 1_{\text Y}$ and Fuller NF cross sections 
agree closely at forward angles, with the 
${\text R} = 1_{\text Y}$ curves providing the correct
continuation of the Fuller NF curves to larger
angles, where the unphysical Fuller NF contributions become 
dominant over the ${\text R} = 1_{\text Y}$ results. 
In addition, the ${\text R} = 1_{\text Y}$ procedure
clearly identifies the oscillatory pattern of the $^{16}$O + $^{16}$O 
full cross sections at $E_{\text {lab}} =$
480 MeV and 704 MeV as an interference between the NF resummed
subamplitudes.

As a final curiosity, we note that a 'suspect' behavior 
appears in the ${\text R} = 1_{\text Y}$ NF 
cross sections for $\alpha + ^{12}$C at $\theta \gtrsim 40^\circ$.
We find that the corresponding ${\text R} = 1_{\text Y}$
NF LIP are $\approx 0$ for $\theta \gtrsim 60^\circ$.
However, the LIP for the full amplitude
is also $\approx 0$ for $\theta \gtrsim 60^\circ$, indicating 
that the full cross section is dominated by contributions from low 
partial waves in this angular region.

We have repeated the calculations,
substituting the WS form factors used in \cite{KHO02},
with symmetrized WS-like form factors defined by
\begin{equation}
f_{\text {sym}}(r,R,d) = \sinh \frac{R}{d} / 
(\cosh \frac{R}{d} + \cosh \frac{r}{d}),
\end{equation}
where $R$ and $d$ are the radius and diffuseness parameters, 
respectively.
We find that the full
cross section decreases for $\theta \gtrsim 60^\circ$ by more than 
5 orders of magnitude, which exceeds the precision limits of our
optical potential computer code.
At forward angles the effect of the substitution is very
small.
This supports the conjecture that the suspect behavior
in the ${\text R} = 1_{\text Y}$ NF and full cross sections 
arises from the 'cusp', which the usual WS potentials have 
at the origin.
The cusp is expected to produce diffraction scattering for 
the low partial waves, which is equally distributed between the N 
and F subamplitudes.
It is this effect that we observe, and
which disappears on removing the cusp.
Similar effects, although masked by the limited range of the scale
along the ordinate, are also observed in the $^{16}$O + $^{16}$O 
collision at $E_{\text {lab}} =$
1120 MeV, as well as tentatively in the ${\text R} = 1_{\text Y}$ 
N $\alpha + ^{40}$Ca cross
section.

%------------------
\section{Conclusions}
\label{Sec:4}

Our new NF resummation procedure clearly improves 
the original Fuller NF method, as is evident 
from the seven examples discussed here, and from those 
presented elsewhere \cite{ANN01d,ANN02a}.
In all these cases, we obtain
NF resummed cross sections that are more slowly varying and less structured
than the Fuller ones.
On the one hand, our results confirm the utility of 
NF methods for gaining insight into the properties of
the subamplitudes responsible for complicated structures
in cross sections. 
On the other hand, they remind us of the empirical origin of NF 
methods, and suggest caution in the interpretation of  
results obtained from NF techniques.
Using different NF methods lets us check what parts of 
the resulting NF subamplitudes
are independent of the particular technique used.
Only properties stable with respect to different NF
methods, can be considered as manifestations of some 
physical phenomenon.

In addition, we have shown that in all NF analyses, it is 
desirable to investigate the behavior of the LAM. 
This quantity is more sensitive to interference effects than
are the NF cross sections, and a null value (or an oscillatory
behavior around zero) of the NF LAM in a certain angular range
may indicate the dominance of an unphysical contribution.

\begin{acknowledgments}
Support of this research by a PRIN MIUR (I) research grant, the
Engineering and Physical Sciences Research Council (UK) and INTAS (EU)
is gratefully acknowledged.
\end{acknowledgments}

%\bibliography{../bib/SC}% Produces the bibliography via BibTeX.

\end{document}